\documentclass[11pt]{article}
\usepackage{epsf}

\title{\bf Gluon production on two centers and the
effective action approach.}
\author{M.A.Braun$^1$, L.N.Lipatov$^{1,2,3}$, M.Yu.Salykin$^1$,
M.I.Vyazovsky$^1$\\
 $^1$St.Petersburg State University, Russia,\\
$^2$ St. Petersburg Nuclear Physics Institute, Gatchina, Russia,\\
$^3$ II Insyitute of Theoretical Physics, Hamburg University, Germany
}
\def\beq{\begin{equation}}
\def\eeq{\end{equation}}
\def\tr{{\rm Tr}\,}
\def\pd{\partial}

\def\lra{\leftrightarrow}

\begin{document}
\maketitle
\medskip

{\bf Abstract}

Application of the effective action formalism is studied for
processes in which the reggeons may split. It is shown that the gluon
production on two centers is described by the contribution of the
Reggeon-to-two-Reggeons-plus-Particle vertex supplemented by certain
singular contributions from the double gluon exchange. The rules for
longitudinal integrations are established from the comparison to
perturbative QCD amplitude. Convenient expressions for application
to the inclusive gluon production are derived.

\section{Introduction}
In the framework of the perturbative QCD, in the Regge kinematics,
particle interaction can be described by the exchange of reggeized gluons
which emit and absorb real gluons and also may split into several
reggeized gluons. Emission of real gluons from  reggeized gluons is
described by effective vertices introduced in  ~\cite{bfkl}
and ~\cite{bartels} for non-split and split reggeons.
Originally both type of vertices were calculated directly
from the relevant simple Feynman diagrams in the Regge kinematics.
Later a powerful effective action formalism was proposed in
~\cite{lipatov}, which considers reggeized and normal gluons as independent
entities from the start and thus allows to calculate all QCD diagrams
in the Regge kinematics automatically and in a systematic and self-consistent
way. However the resulting expressions are 4-dimensional and need
reduction to the final 2-dimensional transverse form. This reduction is
trivial  for tree diagrams but becomes less trivial for diagrams
with loops.

In the paper of two co-authors of the present paper (M.A.B. and M.I.V.)
~\cite{bravyaz} it was demonstrated that the diffractive
amplitude for the production of a real gluon calculated by means of the
effective action and based on the transition of a Reggeon into one or two
Reggeons and Particle (R$\to$R(R)P transition), after integration over
longitudinal variables, coincides with the results calculated in terms
of the above-mentioned effective vertices
(in the purely tranverse "BFKL-Bartels formalism").
However in the process of reduction to the transverse form a certain
prescription had to be used to give sense to divergent integrals.

In this paper we study a more general case: the gluon production
off two different targets. In this case, in the lowest order, with which we
restrict ourselves, the production amplitude by itself
is a tree diagram, without any internal integrations. Longitudinal
integrals appear only in the inclusive cross-section.
Our purpose is twofold. First we analyze application of the
effective action formalism to the processes where the number of
reggeons can be changed (from one to two in our case). As we shall show
this application  requires a careful study of kinematical regions in
which particular diagrams generated by the effective action or
even parts of these diagrams are to be taken into account.
A separate problem
is understanding in which sence  singularities which appear in
these diagrams when the "-" components of the momenta transferred
to the two targets ("energies"), $q_{1,2-}$,  vanish are to be understood.
Comparison with the standard perturbation approach allows
to solve this problem. Note that this latter point was studied in
~\cite{HBL,hentschinski} for simpler diagrams, without reggeon proliferation.

Our second aim is to obtain convenient expressions for the production
amplitudes which can be used for the calculation of the inclusive
cross-sections. The essential region for the integration over transferred
energies involved in this calculation is $q_{1,2-}>>p_-$, where $p_-$
is the energy of the emitted gluon. The amplitude can be drastically simplfied
in this region and transformed to the expression convenient for the following
integrations.

Our results show that in the general kinematics the production amplitude is
almost completely given by the contribution from the R$\to$RRP effective
vertex
derived in ~\cite{bravyaz}. The pole at $q_{1,2-}=0$ should be taken in the
principal values prescription, in agreement with the assumption made in
~\cite{bravyaz}. However the contribution from the R$\to$RRP vertex
should be supplemented by terms proportional to $\delta(q_{1,2-})$
coming from the double reggeon exchange.

\begin{figure}[h]
\leavevmode \centering{\epsfysize=0.2\textheight\epsfbox{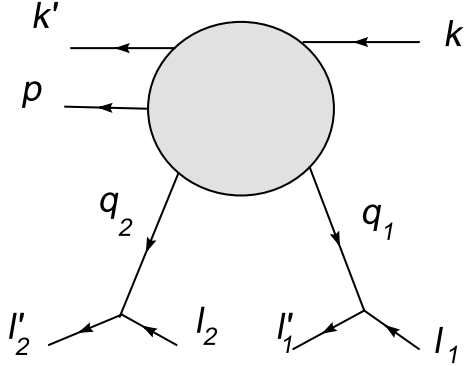}}
\caption{Production amplitude off two scattering centers}
\label{fig1}
\end{figure}

\section{Kinematics and the target lines}
We study the production amplitude on two centers shown in
Fig. \ref{fig1} in the Regge kinematics. For simplicity we consider
the quark as the projectile and scalar quarks as the two scattering centers.
We consider  the central region so that both
$(kp)$ and $(lp)$ are large. This means that $k_+p_-$ and
$l_-p_+$ are much greater than the typical transverse momentum squared
$p_\perp^2$. If $p_+=xk_+$ then our region of $x$ is
\beq
\frac{|p_\perp^2|}{s}<<x<<1.
\eeq
Our longitudinal conservation laws are
\beq
p_++k'_+=k_+,\ \ p_-+q_{1-}+q_{2_-}=0,
\label{conslaw}
\eeq
where we neglected $q_{1+}$, $q_{2+}$ and $k'_-$.

We consider both targets at the same initial momentum $l_1=l_2=l$
in the center of mass system with $l_+=l_\perp=0$.
The transferred momenta in the two collisions are $q_1$ and $q_2$.
The mass-shell conditions give
\beq
2q_1l+q_1^2=2q_2l+q_2^2=0.
\eeq
The initial projectile momentum $k$ has $k_-=k_\perp=0$.
Our Regge kinematics requires that $q_{1-},q_{2-}<<l_-$.
So we put $q_{1-}=\xi_1l_-$ and $q_{2-}=\xi_2l_-$ with
$\xi_{1,2}<<1$. The parameters $\xi_1$ and $\xi_2$ are actually the
conventional Sudakov variables for the transverred momenta $q_1$ and $q_2$.
From
\[q_{1+}=l_{1+}=-\frac{{l'_{1\perp}}^2}{2l'_{1-}}\]
it follows that
\[ q_1^2=2q_{1+}q_{1-}+q_{1\perp}^2=
(1-\xi_1)q_{1\perp}^2\simeq q_{1\perp}^2,\]
so that $q_1$ is almost a purely transverse momentum. as well as $q_2$.

In practical applications we use the axial gauge
in which the gluon field $V$ satisfies $(Vl)=0$
and its propagator is
\beq
P_{\mu\nu}(q)=\frac{h_{\mu\nu}(q)}{q^2},
\ \ h_{\mu\nu}(q)=g_{\mu\nu}-\frac{l_\mu q_\nu+q_\mu l_\nu}{(ql)}.
\eeq
Coupling to the scalar targets generates  vectors
\beq
P_\mu(q)=P_{\mu\nu}(2l^\nu+q^\nu)=P_{\mu\nu}q^\nu=
-l_\mu\frac{q^2}{(ql)}\,\frac{1}{q^2}=2l_\mu\frac{1}{q_\perp^2},
\eeq
where $q=q_1$ or $q=q_2$.
So in fact we can study the amplitude with amplutated target lines
and external legs $P_{\mu_1}(q_1)$ and $P_{\mu_2}(q_2)$.

\section {Gluon emission in the effective action formalism}

In the effective action formalism all real and virtual particles
in the direct channels  split into  groups in correspondence
with their rapidities $y=\frac{1}{2}\ln|p_+/p_-|$.
Gluons with rapidities within some interval $[y-\eta/2, y+\eta/2]$
are described by the usual gluon field $V^{y}_{\mu}=-it^{a}V^{ya}_{\mu}$.
The reggeon field $A^{y}_{{\mu}}=-it^{a}A^{ya}_{\mu}$ with only
non-zero longitudinal components $A_{+}$ and $A_{-}$ corresponds to
virtual gluons in the crossing channels responsible
for the interaction between the groups with essentially different rapidities.

The effective Lagrangian describes the self-interaction of
gluons inside of each group by means of the usual QCD Lagrangian
${\cal L}_{QCD}$ and their interaction with reggeons.
It takes the form \cite{lipatov}:
\beq
{\cal L}_{eff}={\cal L}_{QCD}(V^{y}_\mu+A^{y}_\mu)
+ 2 \tr\Big(({\cal A}_+(V^{y}_+ +A^{y}_+)-A^{y}_+)\pd^2_{\perp} A^{y}_-
+ ({\cal A}_-(V^{y}_- +A^{y}_-)-A^{y}_-)\pd^2_{\perp} A^{y}_+\Big),
\label{e1}
\eeq
where
$$
{\cal A}_{\pm}(V_{\pm})=-\frac{1}{g}\pd_{\pm}\frac{1}{D_{\pm}}\pd_{\pm}*1=
\sum_{n=0}^{\infty}(-g)^nV_{\pm}(\pd_\pm^{-1}V_\pm)^n
$$
\beq
=V_{\pm}-gV_{\pm}\pd_\pm^{-1}V_\pm+g^2V_{\pm}\pd_\pm^{-1}V_\pm
\pd_\pm^{-1}V_\pm+ -...
\label{e2}
\eeq
It is local in rapidity, so the rapidity index $y$ can be omitted.
The shift $V_\mu\to V_\mu+A_\mu$ with $A_\perp=0$ is done
to exclude direct gluon-reggeon transitions.

The reggeon propagator in momentum representation
\beq
<A_+^{y'a}A_-^{yb}>=-i\frac{\delta_{ab}}{q_\perp^2}
\,\theta(y'-y-\eta)
\label{e3}
\eeq
is to be contracted
with field $A_-$ interacting with a group of a higher rapidity $y'$
and field $A_+$ interacting with a group of a smaller rapidity $y$.
From the kinematical constraints it easily follows
that the momentum $q_-$ of  field $A_+$ is small compared to ``-''
components of momenta flowing in the group with a higher rapidity,
so the kinematical condition is implied
\beq
\partial_- A_+ =0.
\label{e4}
\eeq
Analogously, the kinematical condition
\beq
\partial_+ A_- =0
\label{e5}
\eeq
reflects the comparative smallness of the momentum $q_+$
of the field $A_-$.

Formally, in the framework of the effective action approach
one can introduce quite a number of diagrams which describe
interaction with two centers. But one has to take into account
that the rules assumed in the derivation of effective action
put certain restrictions on the kinematical regions appropriate
for particular contributions, so that many of the diagrams which
can formally be introduced have in fact to be dropped once
a particular kinematics is condidered. For some other diagrams
these restrictions may lead to neglecting  some terms.
In short these conditions require that real particles should be emitted
in the multiregge kinematics described above. The emitted particle
should carry nearly all the "+" component of the momentum of the
higher rapidity reggeons and nearly all the "$-$" component of the
lower rapidity reggeons from which it is emitted.

\begin{figure}[h]
\leavevmode \centering{\epsfysize=0.15\textheight\epsfbox{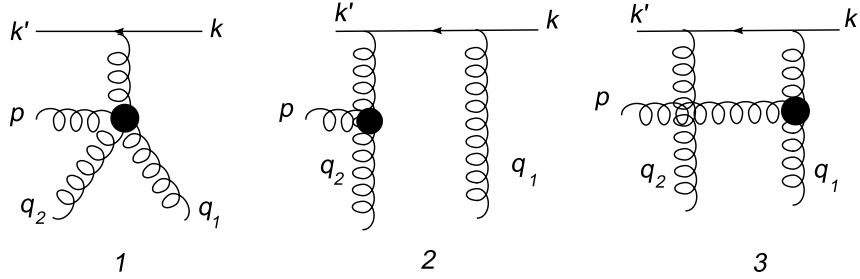}}
\caption{Production amplitude in the effective action formalism}
\label{fig2}
\end{figure}

The essential contribution to the production amplitude
in the effective action formalism is
given by the three diagrams shown in Fig. \ref{fig2}.
Also for the two last diagrams the analogous diagrams with
the interchanged targets ($1\leftrightarrow 2$) have to be added.

The diagram 1 in this figure contains the R$\to$RRP vertex $V$ derived
in ~\cite{bravyaz} in a general kinematics $q_{1,2-} \sim p_-$,
$q_{1,2+} << p_+$:
\beq
V=V_1+V_2,
\eeq
where
\begin{eqnarray}
&& V_1=\frac{i}{2}\,\frac{f^{db_1c}f^{cb_2a}}{(q-q_1)^2+i0}\times
\nonumber \\
&& \Big\{ q_+(4q_1+p)_\mu
-\left((q+q_1)(p-q_2)+q_2^2-q_1^2+(q-q_1)^2
+q_+ q_{1-}\right)n^+_{\mu}
\nonumber \\
&& +\frac{q^2 (q-q_1)^2}{p_- q_{1-}}n^-_{\mu}
+\Big(2q_+ -\frac{q^2}{q_{1-}} \Big)
\Big[-2q_+n^-_{\mu}+(p+2q_2)_\mu+
(p_- -q_{2-}+\frac{q_2^2}{q_+})n^+_{\mu}\Big]\Big\}
\cdot{e^{\mu}}
\label{12a}
\end{eqnarray}
and $V_2=V_1(1\lra 2)$. Here $q=p+q_1+q_2$ with $q^2=q_{\perp}^2$
is the momentum transferred from the projectile,
$e^{\mu}$ is the gluon polarization vector and
$n_{\mu}^{\pm}=(1,0,0,\mp 1)/\sqrt{2}$ are the light-cone unit
\footnote{
Note that the normalization
$a_{\mu}b^{\mu}=a_{+}b_{-}+a_{-}b_{+}+a_{\perp}b_{\perp}$
for the longitudinal components is used here,
whereas it was taken as
$a_{\mu}b^{\mu}=\frac{1}{2}a_{+}b_{-}+\frac{1}{2}a_{-}b_{+}
+a_{\perp}b_{\perp}$ in \cite{bravyaz,braun2}.
}
vectors.
This vertex is found to be transversal with respect to
the gluon momentum $p$.

In the chosen gauge $(Vl)=0$ which is equivalent to $V_+ =0$,
the vertex crucially simplifies:
\begin{equation}
V_1=i\frac{f^{db_1c}f^{cb_2a}}{(q-q_1)^2+i0}
\Big\{2q_+(eq)_{\perp} -\frac{q_\perp^2}{q_{1-}}
\Big[(e,q-q_1)_{\perp}
-\frac{(q-q_1)^2}{p_\perp^2}(ep)_{\perp}\Big]\Big\}\ .
\label{12}
\end{equation}
It is possible to present the contribution (\ref{12}) in
terms of effective R$\to$RP vertices.
In the second term in the brackets
we present the product $1/q_{1-}[(q-q_1)^2+i0]$ as
\[
\frac{1}{q_{1-}[(q-q_1)^2+i0]}
=\frac{2q_+}{(q-q_1)_\perp^2[(q-q_1)^2+i0]}
+\frac{1}{q_{1-}(q-q_1)_\perp^2}
\]
to obtain
\beq
V_1=W_1+R_1 ,
\eeq
where
\beq
W_{1}=-i\frac{2q_+q_{\perp}^2}{(q-q_1)^2+i0}f^{db_1c}f^{cb_2a}
B(p,q_2,q_1),
\eeq
\beq
R_{1}=i\frac{q_{\perp}^2}{q_{1-}}f^{db_1c}f^{cb_2a}
L(p,q_2)
\eeq
and vertices $L$ and $B$ are
\beq
L(p,q_1)=\frac{(p\epsilon_\perp)}{p_\perp^2}-
\frac{(p+q_1,\epsilon_\perp)}{(p+q_1)_\perp^2},\ \
B(p,q_1,q_2)=L(p+q_1,q_2).
\label{lev}
\eeq
The contribution from $R_1$ contains the
singularity at $q_{1}=0$. In  ~\cite{bravyaz} this singularity
was understood in the principal value prescription.
This rule will be proven when we compare the contribution from
the vertex (\ref{12}) to the emission amplitude with the expression
found by the standard perturbation technique.

Coupling to the projectile and taking into account
the upper reggeon propagator
we obtain the part of the amplitude generated by the effective R$\to$RRP
vertex
\beq
{\cal A}^{ef}_1= 32(kl)^2\frac{1}{2k_+q_\perp^2}(V_1+V_2) t^d\, .
\eeq

For the following comparison it is convenient to separate
contributions with different
polarization factors
\[
{\cal A}^{ef}_1=i 32(kl)^2t^d
\Big\{\frac{p_+}{k_+}\frac{(eq)_\perp}{q_\perp^2}
\Big(\frac{f^{db_1c}f^{cb_2a}}{(q-q_1)^2+i0}+
\frac{f^{db_2c}f^{cb_1a}}{(q-q_2)^2+i0}\Big)\]
\[+
\frac{(ep)_\perp}{p_\perp^2}\Big(\frac{f^{db_1c}f^{cb_2a}}{2k_+q_{1-}}+
\frac{f^{db_2c}f^{cb_1a}}{2k_+q_{2-}}\Big)\]
\beq
-(e,q-q_1)_\perp \frac{f^{db_1c}f^{cb_2a}}{2k_+q_{1-}}\frac{1}{(q-q_1)^2+i0}-
(e,q-q_2)_\perp \frac{f^{db_2c}f^{cb_1a}}{2k_+q_{2-}}\frac{1}{(q-q_2)^2+i0}\Big\}
\ .
\label{aeff}
\eeq

For the calculation of the diagrams 2,3 in Fig. \ref{fig2}
one must take into account that with the kinematical conditions
$q_{1,2+}<<p_+ <<k_+$ and $k_+ q_{1,2-} >> q_{1,2\perp}^2,p_{\perp}^2$
the following approximation for the denominators
of the quark propagators can be done
\[
(k-q_1)^2+i0=-2k_+ q_{1-} + q_{1\perp}^2 +i0
\approx -2k_+ q_{1-} +i0\ ,
\]
\beq
(k-p-q_1)^2+i0=2k_+ (-p_{-}-q_{1-}) + (p+q_1)_{\perp}^2 +i0
\approx 2k_+ q_{2-} +i0\ .
\label{prop2}
\eeq
The total contribution from these two diagrams
together with terms $(1\lra 2)$ is given by
\beq
{\cal A}^{ef}_2+{\cal A}^{ef}_3+(1\lra 2)=
\frac{32(kl)^2}{2k_+q_{2-}+i0}L(p,q_1)f^{b_1ac}t^{b_2}t^c+
\frac{32(kl)^2}{-2k_+q_{1-}+i0}L(p,q_2)f^{b_2ac}t^ct^{b_1}+
\Big(1\lra 2\Big).
\label{ampp}
\eeq

Using
$$
\frac{1}{\pm 2k_+q_{1,2-} +i0}
=\pm\frac{1}{2k_+}\cdot P\frac{1}{q_{1,2-}}
-i\pi\delta(2k_+ q_{1,2-})
$$
we separate parts with or without $\delta$-function, which we denote
by upper indeces 0 and 1 respectively.
We find (suppressing factor $32(kl)^2$)
\beq
\Big({\cal A}^{ef}_2+{\cal A}^{ef}_3+(1\lra 2)\Big)^{(0)}=
-i\pi\delta(2k_+q_{2-})L(p,q_1)f^{b_1ac}\{t^{b_2},t^c\}
-i\pi\delta(2k_+q_{1-})L(p,q_2)f^{b_2ac}\{ t^{b_1},t^c\}
\label{ampp0}
\eeq
and
\beq
\Big({\cal A}^{ef}_2+{\cal A}^{ef}_3+(1\lra 2)\Big)^{(1)}=
\frac{L(p,q_1)}{2k_+}f^{b_1ac} [t^{b_2},t^c] P\frac{1}{q_{2-}}+
\frac{L(p,q_2)}{2k_+}f^{b_2ac}[t^{b_1},t^c] P\frac{1}{q_{1-}}\ .
\label{ampp1}
\eeq

However one should take into account the mentioned restriction
to the multiregge character of particle emission.
Then one immediately finds that diagrams 2 and 3 can only contribute in the
region of $q_{1-}$ or $q_{2-}$ close to zero, since otherwise the "$-$' component
of the lower reggeon from which the gluon is emitted will not be totally
transferred to this gluon. This implies that only terms with
$\delta(q_{1.2-})$ should be retained in the contributions from the
diagrams 2 and 3.

This circumstance can be explained in more detail by
the above-mentioned condition of locality in rapidity.
For example, in the diagram 2 the difference between
the rapidity of the projectile
$\frac{1}{2}\ln\frac{s}{M^2}$,
where the small quark mass $M$ is introduced for finiteness,
and the rapidity of the virtual quark with the momentum
$k-q_1$:
\beq
\frac{1}{2}\ln\left|
\frac{k_+ -q_{1+}}{k_- -q_{1-}}
\right|
\label{e6}
\eeq
cannot be more than the cut-off parameter $\eta$.
This condition is equivalent to
\beq
\frac{M^2}{{s}} e^{-\eta} <
\frac{|k_{-}-q_{1-}|}{|k_{+}-q_{1+}|}
< \frac{M^2}{{s}} e^{+\eta} \ .
\label{e7}
\eeq
As pointed out in \cite{lipatov},
the parameter $\eta$ have to be chosen numerically large
but significantly smaller than the relative rapidities
of colliding particles:
$$
1<< \eta << \ln\frac{s}{M^2}\ .
$$
For our kinematical conditions $q_{1+}\to 0$, $k_{-}\to 0$
and the appropriate $\eta$
the inequality (\ref{e7}) leads to the restriction
\beq
|q_{1-}| < \frac{M^2}{\sqrt{s}}e^{+\eta}\ ,
\label{e8}
\eeq
where the limit tends to zero for large $s$.
It follows that the result of the calculation of the diagram
has to be multipied by some $\theta$-function
which allows the values of $q_{1-}$ to be
only within a very small interval around the point
of the pole at $q_{1-}=0$.
For the diagram 3 the same condition reads for $q_{2-}$.
This means that if we understand (\ref{ampp})
in the sense of generalized function
then  terms (\ref{ampp1}) should be dropped
and only the $\delta$-function terms (\ref{ampp0})
remain.

It can be shown that for the diagram 1 in Fig. \ref{fig2}
from the same analysis of rapidity restrictions
it does not follow such a narrow limit for values of $q_{1,2-}$.
Qualitatively, one can say that
the multiperipheral diagram 1 describes emission
for arbitrary relations between $q_{1-}$, $q_{2-}$ and $p_{-}$
excluding the regions close to $q_{1,2-}=0$
but the diagrams 2,3 describe emission from this quasi-elastic
scattering region when $q_{1-}=0$ or $q_{2-}=0$.

As a result, in the effective action formalism the total amplitude
${\cal A}^{ef}$ is given
by the sum of contribution from the R$\to$RRP vertex (\ref{aeff}) and
the part (\ref{ampp0}) of the double reggeon exchange containing
$\delta(q_{1,2-})$. In the following section we shall see that this
identically coincides with the amplitude calculated by the standard
perturbative QCD provided the singularities at $q_{1,2-}=0$ in
the part with R$\to$RRP vertex are understood in the principal value
sense.

\begin{figure}[h]
\leavevmode \centering{\epsfysize=0.15\textheight\epsfbox{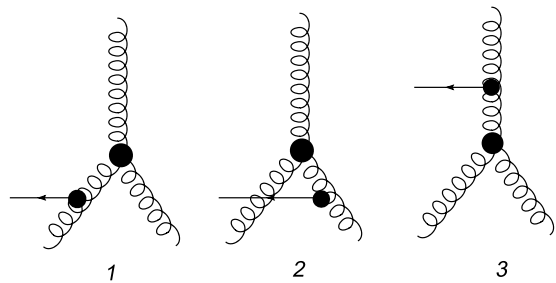}}
\caption{Diagrams containing the three-reggeon vertex}
\label{fig3}
\end{figure}

In the effective vertex formalism
the three diagrams in Fig. \ref{fig3} containing the R$\to$RR
transition (three-reggeon) vertex can also be drawn.
The found contribution from the diagram in Fig. \ref{fig2},1
actually transforms into the contribution of one
of these diagrams depending on the correct multiregge character
of emission in a given kinematics. Therefore
these diagrams should not be added to the amplitude ${\cal A}^{ef}$
to avoid double counting.
They rather describe emisssion for particular relations
between $p_-,\,q_{1-}$ and $q_{2-}$.
For example the diagram in Fig. \ref{fig3},2
corresponds to the situation when $p_-\simeq -q_{1-}>>q_{2-}$.
Of special interest is the diagram
in Fig. \ref{fig3},3, which corresponds to the situation when
$|q_{1,2-}|>>p_-$ considered in Section 5.

\begin{figure}[h]
\leavevmode \centering{\epsfysize=0.15\textheight\epsfbox{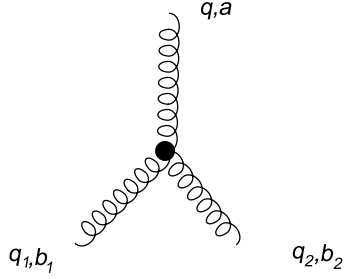}}
\caption{The vertex R$\to$RR}
\label{fig4}
\end{figure}

The three-reggeon vertex is determined by the
contribution $S_3$ to the action ~\cite{lipatov}. The relevant term
in the Lagrangian is
\beq
{\cal L}_3=
-2g{\rm Tr}\,\Big( A_-\partial^{-1}_-A_-\partial^2_\perp A_+\Big)
=-{g}{\rm Tr}\,\Big(\Big[ A_-,\partial^{-1}_-A_-\Big]
\partial^2_\perp A_+\Big).
\eeq
It corresponds to the vertex shown in Fig. \ref{fig4}:
\[
{\cal L}_3=-{g}(-i)^3A_-^{b_1}\partial_-^{-1}A_-^{b_2}
\partial^2_\perp A_+^a{\rm Tr}\Big([t^{b_1},t^{b_2}] t^a\Big) \, .
\]
We have
\[
{\rm Tr}\Big([t^{b_1},t^{b_2}] t^a\Big)=\frac{i}{2}f^{ab_1b_2}
\]
and also
\[\partial_-^{-1}A_-^{b_2}=-i\frac{1}{q_{2-}}A_-^{b_2},\ \
\partial^2_\perp A_+^a=-q_{\perp}^2A_+^a \ . \]
It leads to the vertex
\beq
\Gamma_{R\to RR}=\frac{q_\perp^2}{2q_{1-}}f^{ab_1b_2}=
-\frac{q_\perp^2}{2q_{2-}}f^{ab_1b_2} \ .
\label{rrrv}
\eeq
Note that the kinematical condition $\partial_- A_+ =0$
applied to ${\cal L}_3$ is equivalent to
the momentum relation $q_{1-}+q_{2-}=0$
which is fully correspondent to
the considered case $|q_{1,2-}|>>p_-=-(q_{1-}+q_{2-})$.

Using the vertex we find that the contribution of diagram 3 in
Fig. \ref{fig3} to the amplitude is
\beq
{\cal A}^{ef}_{R\to RR}= 32(kl)^2\frac{1}{2k_+q^2}t^d V_3 \, ,
\eeq
where
\beq
V_3=
i\frac{q_\perp^2}{q_{1-}}L(p,q_1+q_2)f^{adc}f^{cb_1b_2} \ .
\label{add2}
\eeq
As it will be calculated in Section 5, this contribution
is exactly equal to the expression
(\ref{aa}) obtained for the antisymmetric part
of the amplitude in the kinematics $q_{1,2-}>>p_-$.
Clearly, it also coincides with the contribution of the diagram
of Fig. ~\ref{fig2},1 with the R$\to$RRP vertex in this limit,
since the latter gives the part of the amplitude
free from $\delta(q_{1,2})$ terms in the general case.
It assumes the principal value prescription also
for the pole at $q_{1,2-}=0$ in the vertex (\ref{rrrv}).
So in this particular kinematics the part
of the amplitude without $\delta$-functions
can be fully described by the diagram 3 in Fig. \ref{fig3}.

\section{Production amplitude in the lowest order of the perturbative QCD}

\begin{figure}[h]
\leavevmode \centering{\epsfysize=0.35\textheight\epsfbox{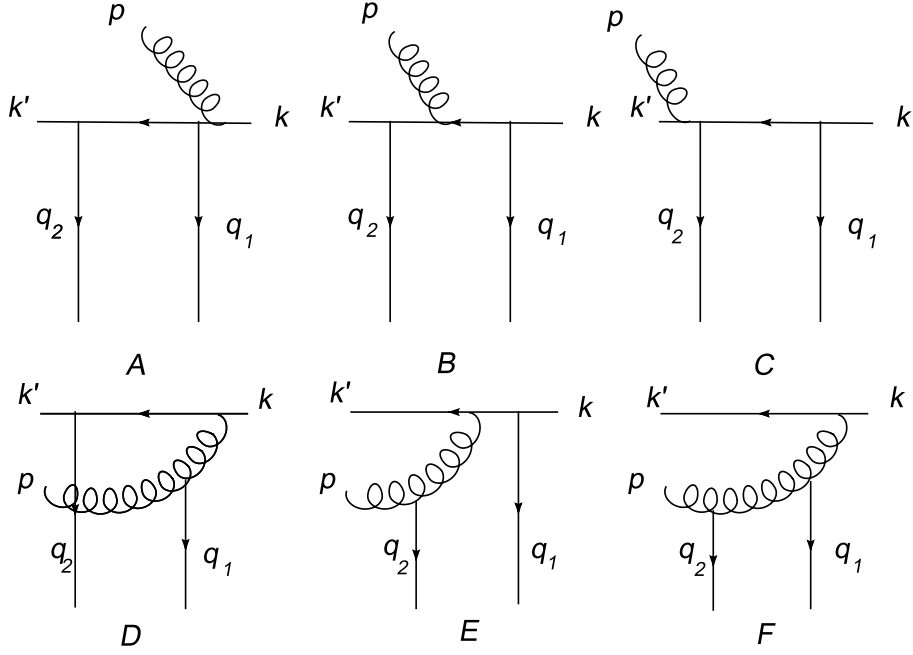}}
\caption{Production amplitude off two scattering centers in the
lowest order}
\label{fig5}
\end{figure}

We study production of a gluon with momentum $p$ in collision
of the projectile with momentum $k$ and two targets with
their intial momenta $l$ each.
In the standard Feynman formalism the production amplitude
is given by 6 diagrams $A,...,F$ in Fig. \ref{fig5}
summed with the same diagrams with the transposed targets
$1\leftrightarrow 2$.

\subsection{Diagrams $A,B,C$}
We start from three similar diagrams $A,B$ and $C$.
In all the three the quark line generates a momentum factor
\beq
M=-32i(kl)^2\frac{k_+}{p_+}(ep)_\perp.
\eeq
Here
the factor $-i$ comes from 2 vertices, 2 propagators and the minus sign
which originates from the relation
\beq
(ek)=e_-k_+=-\frac{k_+}{p_+}(ep)_\perp.
\eeq
The colour factors are
\beq
C_A=t^{b_2}t^{b_1}t^a,\ \ C_B=t^{b_2}t^{a}t^{b_1},\ \
C_C=t^at^{b_2}t^{b_1}.
\eeq
The propagators contribute factors
\[
P_A=\frac{1}{[(k-p)^2+i0][(k'+q_2)^2+i0]},\ \
P_B=\frac{1}{[(k-q_1)^2+i0][(k'+q_2)^2+i0]},\]\beq
P_C=\frac{1}{[(k-q_1)^2+i0][(k'+p)^2+i0]}.
\eeq
Using $k_+p_->>|p_\perp^2|$ we can write
\[(k-p)^2=-(k'+p)^2=\frac{k_+}{p_+}p_\perp^2\]
to finally obtain
\beq
A=-32i(kl)^2\frac{(ep)_\perp}{p_\perp^2}
\frac{1}{(k'+q_2)^2+i0}t^{b_2}t^{b_1}t^a
\eeq
and
\beq
C=32i(kl)^2\frac{(ep)_\perp}{p_\perp^2}
\frac{1}{(k-q_1)^2+i0}t^at^{b_2}t^{b_1}.
\eeq

As to the contribution $B$ we can formally write it in a similar
fashion
\beq
B=-32i(kl)^2\frac{(ep)_\perp}{p_\perp^2}\frac{k_+}{p_+}
\frac{p_\perp^2}{[(k-q_1)^2+i0][(k'+q_2)^2+i0]}t^{b_2}t^at^{b_1}.
\eeq
Its relative weight depends on the values of $q_{1-}$ and $q_{2-}$.
They are constrained by the conservation law (\ref{conslaw}), so that their
sum is of the order $p_-$.

\subsection{Diagrams $D$ and $E$}
\begin{figure}[h]
\leavevmode \centering{\epsfysize=0.12\textheight\epsfbox{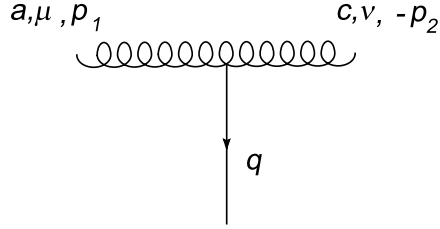}}
\caption{Insertion of an interaction with the target into the gluon line.
All lines are assumed outgoing; $(p_1l)=(p_2l)\equiv(pl)$.}
\label{fig6}
\end{figure}

To study the diagrams $D$, $E$ and $F$ we shall use some properties
of the gluon propagator in the axial gauge, derived in ~\cite{bra4}.
Namely, interaction with the projectile introduces into the gluon line
the vertex (see Fig. \ref{fig6})
\beq
2(pl)g_{\mu\nu}f^{abc}.
\eeq
So the two gluon propagators connected by this vertex, apart from the
standard denominators, contains the momentum factor
\beq
H_{\mu\nu}(p_1,p_2)=g_{\mu\nu}-\frac{l_\mu p_{1\nu}+p_{2\mu}l_\nu}{(pl)}
+\frac{(p_1p_2)l_\mu l_\nu}{(pl)^2}.
\label{hh}
\eeq
Here it is used that the "$+$" component of the gluon momentum does not change
in its interaction with the target.

Coupling to the gluon polarization vector gives
\beq
h_{\mu\nu}(p)e^\nu=e_\mu-l\mu\frac{(ep)}{(pl)}\equiv E_\mu(p),
\eeq
\beq
H_{\mu\nu}(p_1,p_2)e^\nu=E_\mu(p_2).
\label{hhe}
\eeq
Note that we have
\beq
E_-(p)=-\frac{1}{p_+}(ep)_\perp.
\eeq

Armed with these relations we start from  diagram  $D$.
The momentum factor from the quark and gluon lines is
\beq
M_D=32(pl)(kl)\frac{k_+}{p_+}(e,p+q_1)_\perp
\eeq
(we have $i^3$ from two vertices and the propagator in the quark line,
an $(-i)$ from the gluon line and  a $(-1)$ from $E$).
The color factor is
\beq
C_D=f^{ab_1c}t^{b_2}t^c.
\eeq
The two propagators give
\beq
P=\frac{1}{[(k'+q_2)^2+i0][(p+q_1)^2+i0]}.
\eeq
So diagram $D$ gives a contribution
\beq
D=32(kl)^2\frac{(e,p+q_1)_\perp}{(p+q_1)_\perp^2}
\frac{(p+q_1)_\perp^2}{[(k'+q_2)^2+i0][(p+q_1)^2+i0]}
f^{ab_1c}t^{b_2}t^c,
\label{diad}
\eeq
where we used $(pl)=(kl)p_+/k_+$.

Analogous calculations give for diagram $E$
\beq
E=32(kl)^2\frac{(e,p+q_2)_\perp}{(p+q_2)_\perp^2}
\frac{(p+q_2)_\perp^2}{[(k-q_1)^2+i0][(p+q_2)^2+i0]}
f^{ab_2c}t^ct^{b_1}.
\label{eiad}
\eeq

\subsection{Diagram $F$}

Using our formulas (\ref{hh}) and (\ref{hhe}) we get the
momentum factor
\beq
M_F=-32i(pl)^2[kE(p+q_1+q_2)]=
32i(kl)^2\frac{k_+}{p_+}(e,p+q_1+q_2)_{\perp}.
\eeq
The colour factor is
\beq
C_F=f^{ab_2c}f^{cb_1d}t^d.
\eeq
The propagators give
\beq
P_F=\frac{1}{[(p+q_2)^2+i0][(p+q_2+q_1)^2+i0]}.
\eeq
In fact $p_-+q_{2-}+q_{1-}\simeq 0$ so in the second denominator
only the transversal part remains.

We get our final result
\beq
F=32i(kl)^2\frac{p_+}{k_+}\frac{(e,p+q_1+q_2)_{\perp}}{(p+q_1+q_2)_{\perp}^2}
\frac{1}{[(p+q_2)^2+i0]}f^{ab_2c}f^{cb_1d}t^d\ .
\eeq

\subsection{Parts with and without $\delta_(q_{1,2-})$}
To compare with the results of the effective action formalism,
we split the total perturbative contribution into parts with and
without $\delta_(q_{1,2-})$.
We start by rewriting the propagator
in the contribution of the diagram $B$ in Fig. \ref{fig5}
as
\[
P_B=\frac{1}{(-2k_+q_{1-}+i0)(2k_+q_{2-}+i0)}=
\frac{1}{2k_+(q_{1-}+q_{2-})}\Big(\frac{1}{-2k_+q_{1-}+i0}-
\frac{1}{2k_+q_{2-}+i0}\Big)\]\beq=
\frac{1}{2k_+p_-}\Big(\frac{1}{2k_+q_{1-}-i0}+
\frac{1}{2k_+q_{2-}+i0}\Big).
\eeq
So we find
\beq
B=32i(kl)^2\frac{(ep)_\perp}{p_\perp^2}
\Big(\frac{1}{2k_+q_{1-}-i0}+\frac{1}{2k_+q_{2-}+i0}\Big)
t^{b_2}t^at^{b_1}.
\eeq

Combining this contribution with terms $A$ and $C$
we obtain
\beq
A+B+C=-i32(kl)^2\frac{(ep)_\perp}{p_\perp^2}
\Big\{\frac{t^{b_2}[t^{b_1},t^a]}{2k_+q_{2-}+i0}+
\frac{[t^a,t^{b_2}]t^{b_1}}{2k_+q_{1-}-i0}\Big\}.
\label{abc}
\eeq

The contribution with 1 and 2 interchanged is
\beq
(A+B+C)(1\lra 2)=
-i32(kl)^2\frac{(ep)_\perp}{p_\perp^2}
\Big\{\frac{t^{b_1}[t^{b_2},t^a]}{2k_+q_{1-}+i0}+
\frac{[t^a,t^{b_1}]t^{b_2}}{2k_+q_{2-}-i0}\Big\}.
\label{abc1}
\eeq

Separating parts with and without the  $\delta$ function,
which we again denote by upper indices $0$ and $1$ respectively we obtain
\[
\Big(A+B+C+(1\lra 2)\Big)^{(0)}=
-i32(kl)^2\pi \frac{(ep)_\perp}{p_\perp^2}
\Big(\delta(2k_+q_{2-})f^{b_1ac}\{t^{b_2},t^c\}+
\delta(2k_+q_{1-})f^{b_2ac}\{t^{b_1},t^c\}\Big)
\]
and
\[
\Big(A+B+C+(1\lra 2)\Big)^{(1)}=
i32(kl)^2\frac{(ep)_\perp}{p_\perp^2}t^d\Big(
\frac{P}{2k_+q_{2-}}f^{b_2cd}f^{b_1ac}+
\frac{P}{2k_+q_{1-}}f^{b_1cd}f^{b_2ac}\Big).
\]

We pass to our diagrams Fig. \ref{fig5} $D$ and $E$.
We rewrite our formulas (\ref{diad}) and (\ref{eiad}) as
\[
D=32(kl)^2(e,p+q_1)_\perp\Big(\frac{P}{2k_+q_{2-}}-
i\pi\delta(2k_+q_{2-})\Big)
\frac{1}{(p+q_1)^2+i0}
f^{ab_1c}t^{b_2}t^c
\]
and
\[
E=-32(kl)^2(e,p+q_2)_\perp\Big(\frac{P}{2k_+q_{1-}}+i\pi
\delta(2k_+q_{1-})\Big)\frac{1}{(p+q_2)^2+i0}
f^{ab_1c}t^{b_2}t^c.
\]
Adding the terms with $1\lra 2$ we find the principal value and
$\delta$-function parts of the sum as
\[
\Big(D+E+(1\lra 2)\Big)^{(0)}\]\[=
 -i32(kl)^2\pi\frac{(e,p+q_1)_\perp}{(p+q_1)_\perp^2}
\delta(2k_+q_{2-})f^{ab_1c}\{t^{b_2},t^c\}
 -i32(kl)^2\pi\frac{(e,p+q_2)_\perp}{(p+q_2)_\perp^2}
\delta(2k_+q_{1-})f^{ab_2c}\{t^{b_1},t^c\}
\]
and
\[
\Big(D+E+(1\lra 2)\Big)^{(1)}\]\[=
 i32(kl)^2 \frac{P}{2k_+q_{2-}}\frac{(e,p+q_1)_\perp}{(p+q_1)^2+i0}
f^{b_2cd}f^{ab_1c}t^d
 +i32(kl)^2\frac{P}{2k_+q_{1-}}\frac{(e,p+q_2)_\perp }{(p+q_2)^2+i0}
f^{b_1cd}f^{ab_2c}t^d.
\]
In the derivation of the part with the $\delta$-function we used that
\[\delta(2k_+q_{2-})\frac{1}{(p+q_1)^2+i0}=
\delta(2k_+q_{2-})\frac{1}{(p+q_1)^2_\perp},\]
since in this relation $(p+q_1)_-=-q_{2-}=0$.

In the general kinematics  diagram $F$ does not contain denominators
singular in $q_{1,2-}$. So the only contribution is
\[\Big(F+(1\lra 2)\Big)^{(1)}=
i 32(kl)^2\frac{p_+}{k_+}\frac{(e,p+q_1+q_2)_\perp}{(p+q_1+q_2)_\perp^2}t^d
\Big(\frac{f^{ab_2c}f^{cb_1d}}{(p+q_2)^2+i0}+
\frac{f^{ab_1c}f^{cb_2d}}{(p+q_1)^2+i0}\Big).
\]

Comparison of the total perturbative amplitude demonstrates that
the terms without the $\delta(q_{1,2-})$ sum into exacttly the
expression (\ref{aeff}) coming from the R$\to$RRP vertex, provided we
interprete
in the latter the singularities at $q_{1,2-}=0$ in the principal value
prescription. The terms
containing  $\delta(q_{1,2-})$ are exactly reproduced by  the corresponding
part (\ref{ampp0})
of the contribution from the double gluon exchange. Thus comparison
of the amplitudes
obtained in the effective action
formalism and perturbative QCD demonstrates that they are completely
identical. As a byproduct of this comparison we find that the singularties
at $q_{1,2-}=0$ in the effective R$\to$RRP vertex should be understood
in the principal value prescription.

\section{Production amplitude in the region $q_{1-},q_{2-}>>p_-$}

The "-"-components of momenta $q_{1,2}$ transferred to the targets
can be taken arbitrary. In fact in the inclusive cross-sections one
integrates over all values of $q_{1-}$ and $q_{2-}$ related by the
conservation law (\ref{conslaw}). The two production amplitudes enter
into the inclusive cross-section at different momenta of target quarks,
shifted by the momentum $\lambda$ transferred to the nucleus.
Its value is determined by the properties of the nucleus. In the rest
system of the target $\lambda_-\sim\sqrt{m\epsilon}$ where
$m$ is the nucleon mass and $\epsilon$ is the binding energy. In the same
system in the central emission region $p_-\sim |p_\perp|m/\sqrt{s}$, so that
$p_-<<\lambda_-$. As a result, in the integration over $q_{1,2-}$
the essential values of $q_{1,2-}$ are determined by $\lambda_-$ and so
are much larger than $p_-$
In this kinematical region
\beq
q_{1-},q_{2-}>>p_-,\ \ q_{-}+q_{2-}=0
\label{cond}
\eeq
our expressions for the amplitude can be substantially simplified.

With (\ref{cond}) we have
\[
\Big|\frac{k_+}{p_+}\frac{1}{(k-q_1)^2}\Big|
=\Big|\frac{1}{p_+q_{1-}}\Big|<<\Big|\frac{1}{p_+p_{-}}\Big|=
\Big|\frac{1}{p_\perp^2}\Big|
\]
and the contribution $B$ is much smaller than $A$ and $C$. So in the
kinematics (\ref{cond}) the diagram $B$ can be neglected.
Also with (\ref{cond}) we have
\beq
(k-q_1)^2=(k'+q_2)^2=2k_+q_{2-}
\eeq
and for the sum $A+C$ we have
\[
A+C=-32i(kl)^2\frac{(ep)_\perp}{p_\perp^2}
\frac{1}{2k_+q_{2-}+i0}(t^{b_2}t^{b_1}t^a-t^at^{b_2}t^{b_1})\]\beq=
32(kl)^2\frac{(ep)_\perp}{p_\perp^2}
\frac{1}{2k_+q_{2-}+i0}(f^{b_1ac}t^{b_2}t^c+f^{b_2ac}t^ct^{b_1}).
\label{tac1}
\eeq

Next in the kinematics (\ref{cond}) we find that the contributions of diagrams
$D$ and $E$ are small, since in them $q_{2-}$ appears squared
in the denominators. However
the contribution from diagram $D$  contains a $\delta$-function, since the
poles in, say, $q_{2-}$ are located on different sides of the real axis.
In our kinematics $q_{1-}=-q_{2-}$ so that
\beq
D=32(pl)^2\frac{(e,p+q_1)_\perp}{(p+q_1)_\perp^2}
\frac{(p+q_1)_\perp^2}{[2k_+q_{2-}+i0][-2p_+q_{2-}+(p+q_1)_\perp^2+i0]}
f^{ab_1c}t^{b_2}t^c.
\label{diad1}
\eeq
Integration over $q_{2-}$ shows that
that $D$ contains a $\delta$-contribution
\beq
\Delta D=-32(kl)^22\pi i\delta(2k_+q_{2-})
\frac{(e,p+q_1)_\perp}{(p+q_1)_\perp^2}
f^{ab_1c}t^{b_2}t^c.
\eeq
In $E$ the two poles in $q_{2-}$ are located on the same side of the real
axis, so that it does not contain $\delta$-like contributions.
So  the only contribution which remains from  $D$ and $E$ in our
kinematics is
\[
D+D(1\lra 2)=\Delta D+\Delta D(1\lra 2)\]\beq=
32(kl)^22\pi i\delta(2k_+q_{1-})\Big(\frac{(e,p+q_2)_\perp}{(p+q_2)_\perp^2}
f^{ab_2c}t^ct^{b_1}+\frac{(e,p+q_1)_\perp}{(p+q_1)_\perp^2}
f^{ab_1c}t^ct^{b_2}\Big).
\eeq

Finally the expression for $E$  can be simplified to
\beq
F=32i(kl)^2\frac{(e,p+q_1+q_2)_{\perp}}{(p+q_1+q_2)_{\perp}^2}
\frac{1}{2k_+q_{2-}+i0}f^{ab_2c}f^{cb_1d}t^d.
\eeq

The total contribution  to the amplitude
\beq
{\cal A}=A+C+D+F+(1\lra 2)
\eeq
can be rewritten in the form which follows the rules
of the BFKL-Bartels formalism. In the following we suppress the
common factor $32(kl)^2$. Having this in mind
we add to our
contribution a term
\beq
T=T^{(1)}+T^{(2)},
\eeq
where
\beq
T^{(1)}=-\frac{1}{2k_+q_{2-}+i0}
\Big(\frac{(e,p+q_2)_\perp}{(p+q_2)_\perp^2}f^{acb_2}t^{b_1}t^c+
\frac{(e,p+q_1)_\perp}{(p+q_1)_\perp^2}f^{acb_1}t^{b_2}t^c\Big)
+\Big(1\lra 2\Big)
\label {term1}
\eeq
and
\beq
T^{(2)}=-2\pi i\delta(2k_+q_{2-})
\Big(\frac{(e,p+q_2)_\perp}{(p+q_2)_\perp^2}f^{acb_2}t^{b_1}t^c+
\frac{(e,p+q_1)_\perp}{(p+q_1)_\perp^2}f^{acb_1}t^{b_2}t^c\Big),
\label{term2}
\eeq
which is zero, since with $q_{1-}=-q_{2-}$
\[\frac{1}{2k_+q_{2-}+i0}+\frac{1}{2k_+q_{1-}+i0}
=-2\pi i\delta(2k_+q_{2-}).\]
We see that term $T^{(2)}$ cancels the contribution
from $D$:
\beq
D+D(1\lra 2)+T^{(2)}=0.
\eeq

We present the remaining  expressions as:
\beq
A+C+(1\lra2)=\frac{(ep)_\perp}{p_\perp^2}\frac{1}{2k_+q_{2-}+i0}
\Big(f^{b_1ac}t^{b_2}t^c+
f^{b_2ac}t^ct^{b_1}\Big)+\Big(1\lra 2\Big)
\label{tac}
\eeq
and
\beq
F+(1\lra 2)=i\frac{(e,p+q_1+q_2)_{\perp}}{(p+q_1+q_2)_{\perp}^2}
\frac{1}{2k_+q_{2-}+i0}
f^{ab_2c}f^{cb_1d}t^d+\Big(1\lra 2\Big)
\label{tf}
\eeq
and shall transform the explicitly shown expressions.

We combine the second term in (\ref{term1}) with the first term in
(\ref{tac}) to get
\beq
T_1=\frac{1}{2k_+q_{2-}+i0}\Big(\frac{(ep)_\perp}{p_\perp^2}-
\frac{(e,p+q_1)_\perp}{(p+q_1)_\perp^2}\Big)f^{b_1ac}t^{b_2}t^c.
\label{t1}
\eeq
We transform the colour factor in the first term in (\ref{term1})
as
\beq
f^{acb_2}t^{b_1}t^c= f^{acb_2}t^ct^{b_1}+if^{acb_2}f^{b_1cd}t^d
\label{rel}
\eeq
and the contribution from the first term in this relation
combine with the second term in (\ref{tac}) to get
\beq
T_2=\frac{1}{2k_+q_{2-}+i0}\Big(\frac{(ep)_\perp}{p_\perp^2}-
\frac{(e,p+q_2)_\perp}{(p+q_2)_\perp^2}\Big)f^{b_2ac}t^ct^{b_1}.
\label{t2}
\eeq
Finally we combine the contribution from the second term in (\ref{rel})
with $F$ and obtain
\beq
T_3=-i\frac{1}{2k_+q_{2-}+i0}\Big(\frac{(e,p+q_2)_\perp}{(p+q_2)_\perp^2}-
\frac{(e,p+q_1+q_2)_{\perp}}{(p+q_1+q_2)_{\perp}^2}\Big)
f^{ab_2c}f^{cb_1d}t^d.
\label{t3}
\eeq

As we observe that all the contributions nicely arrange into  three
terms with R$\to$RP effective vertices.
We find that the total amplitude is a sum
\beq
{\cal A}=\Big({\cal A}_1+{\cal A}_2+{\cal A}_3\Big)+\Big(1\lra 2\Big),
\eeq
where
\beq
{\cal A}_1=\frac{1}{(k'+q_2)^2+i0}L(p,q_1)f^{b_1ac}t^{b_2}t^c,
\label{amp1}
\eeq
\beq
{\cal A}_2=\frac{1}{(k-q_1)^2+i0}L(p,q_2)f^{b_2ac}t^ct^{b_1},
\label{amp2}
\eeq
\beq
{\cal A}_3=-i\frac{1}{2k_+q_{2-}+i0}
B(p,q_2,q_1)f^{ab_2c}f^{cb_1d}t^d=
-i\frac{(ql)}{(kl)}\frac{1}{(p+q_2)^2+i0}B(p,q_2,q_1)f^{ab_2c}f^{cb_1d}t^d.
\label{amp3}
\eeq

\begin{figure}[h]
\leavevmode \centering{\epsfysize=0.15\textheight\epsfbox{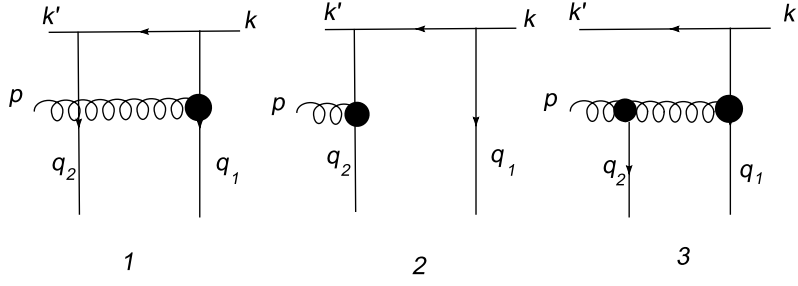}}
\caption{Production amplitude off two scattering centers in terms of the
$L$ and $B$  vertices}
\label{fig7}
\end{figure}

The three terms ${\cal A}_{1,2,3}$ correspond to the diagrams shown in Fig.
\ref{fig7} with the  vertices $L$ and $B$ and expected dependence
on $q_{1-}$ and $q_{2-}$.

The denominator $2k_+q_{2-}+i0$ splits into parts symmetric and antisymmetric
with repect to the change $q_{2-}\to q_{1-}$, that is $u\to s$
\beq
\frac{1}{2k_+q_{2-}+i0}=P\frac{1}{2k_+q_{2-}}-\pi i\delta(2k_+q_{2-}).
\eeq
The first term corresponds to the antisymmetric and the second to
symmetric parts. Note that the $\delta$ term does not literally mean
that $q_{2-}=0$, which seems to violate the adopted kinematics
$q_{2-}>>p_{2-}$. Rather it means that in the integration over $q_{2-}$
in the complex plane one can neglect terms of the order $p_-$ in the integrand
and take the integral around the resulting singularity at $q_{2-}=0$.

It is instructive to find the final expressions for the
parts of the amplitude symmetric and
antisymetric under the interchange $s\lra u$,
$\cal{A}^{(+)}$ and
$\cal{A}^{(-)}$ respectively.  The antisymmetric part is
\[
{\cal{A}}^{(-)}=P\frac{1}{2k_+q_{2-}}\Big[L(p,q_1)f^{b_1ac}t^{b_2}t^c
+L(p.q_2)f^{b_2ac}t^ct^{b_1}-iB(p,q_2,q_1)f^{ab_2c}f^{cb_1d}t^d\]\[-
L(p,q_2)f^{b_2ac}t^{b_1}t^c
-L(p.q_1)f^{b_1ac}t^ct^{b_2}+iB(p,q_1,q_2)f^{ab_1c}f^{cb_2d}t^d\Big].
\]
Combining terms with the same R$\to$RP vertices
and using
\beq
L(p,q_1)+B(p,q_1,q_2)=L(p,q_2)+B(p,q_2,q_1)=L(p,q_1+q_2)
\eeq
we find that the antisymmetric part amplitude is given by a
simple expresssion
\beq
{\cal{A}}^{(-)}=P\frac{1}{2k_+q_{2-}}L(p,q_1+q_2)it^d
(f^{b_1ac}f^{b_2cd}-f^{b_2ac}f^{b_1cd}).
\eeq
Using further the Jacobi identity
\[f^{b_1ac}f^{b_2cd}-f^{b_2ac}f^{b_1cd}=f^{dac}f^{cb_1b_2}\]
we finally obtain
\beq
{\cal{A}}^{(-)}=P\frac{1}{2k_+q_{2-}}L(p,q_1+q_2)it^d
f^{dac}f^{cb_1b_2}.
\label{aa}
\eeq
As expected, in the adopted kinematics
the antisymmetric part gives a contribution
which corresponds to the reggeon diagram 3 (\ref{add2})
in Fig. \ref{fig3}.
From its structure one concludes that it corresponds to the
interaction with the target quarks having the $t$-channel with the
gluon colour and gives no contribution to to the interaction with the
vacuum channel.

The symmetric of the amplitude is given by
\[
{\cal{A}}^{(+)}=-\pi i\delta(2k_+q_{2-})\Big[L(p,q_1)f^{b_1ac}\{t^{b_2},t^c\}
+L(p.q_2)f^{b_2ac}\{t^{b_1},t^c\}\]\[-B(p,q_2,q_1)f^{b_2ac}[t^{b_1},t^c]
-B(p,q_1,q_2)f^{b_1ac}[t^{b_2},t^c]\Big],\]
or, in terms of R$\to$RP vertices,
\[
{\cal A}^{(+)}=-\pi i\delta(2k_+q_{2-})\Big[2L(p,q_1)f^{b_1ac}t^ct^{b_2}
+2L(p.q_2)f^{b_2ac}t^ct^{b_1}-iL(p,q_1+q_2)\Big(f^{b_2ac}f^{cb_1d}+
f^{b_1ac}f^{cb_2d}\Big)t^d\Big]\ .
\]

\section{Conclusions}
We have found that in the general kinematics, out of all possible
diagrams which one can formally draw in the effective action approach,
only quite a few are to be taken into account in accordance with the
requirement of the multiregge kinematics. The contribution from the
R$\to$RRP effective vertex gives all the terms in the amplitude which
do not contain $\delta(q_{1,2-})$. The latter are supplied by the
double reggeon exchange. All the rest diagrams just reproduce the limiting
cases of the R$\to$RRP contribution in different kinematical regions,
that is  relations between $q_{1,2-}$ and $p_-$.

In the kinematics
$q_{1,2-}>>p_-$ appropriate for the calculation of the inclusive
cross-section of gluon production on two centers, the production amplitude
is reproduced by a set of reggeon diagrams with
effective vertices multiplied by energetic factors $1/s$ and $1/u$.
The set is the same as used in the calculation of the inclusive cross-sections
directly in the purely transversal technique. So one expects that after
integration over the intermediate target momenta one will obtain the same
results for the inclusive cross-section as currently obtained in the
purely transversal approaches (BFKL-Bartels or dipole).

A byproduct of our study is justification of the rule of
integration over the singularities at $q_{1,2-}=0$ in the principal value
prescription imposed {\it ad hoc} in ~\cite{bravyaz}. As follows from our
study, terms in the effective vertex with this singularity contribute
only to the part of the amplitude without $\delta$ functions in
the transferred energies
and so should be integrated in this prescription.

\section{Acknowledgement}

The authors are most thankful to J.Bartels and G.P.Vacca for helpful and
constructive discussions. M.A.B. is indebted to the 2nd Institute
of theoretical Physics at Hamburg university
for hospitality and financial support.


\begin{thebibliography}{99}
%
\bibitem{bfkl} L.N.Lipatov, Sov. J. Nucl. Phys. {\bf 23} (1976) 338;
E.A.Kuraev, L.N.Lipatov and V.S.Fadin, Sov. Phys. JETP {\bf 45} (1977) 199;
I.I.Balitsky amd L.N.Lipatov, Sov. J. Nucl. Phys. {\bf 28} (1978) 822.
%
\bibitem{bartels} J.Bartels, Nucl. Phys. {\bf B175} (1980) 365.
%
\bibitem{lipatov} L.N.Lipatov, Nucl. Phys. {\bf B 452} (1995) 369;
 L.N.Lipatov, Phys. Rep. {\bf 286} (1997) 131.
%
\bibitem{bravyaz} M.A.Braun, M.I.Vyazovsky,
Eur. Phys. J. {\bf C 51} (2007) 103.
%
\bibitem{HBL} M.Hentschinski, J.Bartels and L.N.Lipatov, arXiv: 0809.4146.
%
\bibitem{hentschinski} M.Hentschinski, Acta Phys. Polon. {\bf B 39} (2008)
2567; arXiv: 0908.2576.
%
\bibitem{braun2} M.A.Braun, M.Yu.Salykin, M.I.Vyazovsky,
Eur. Phys. J. {\bf C 65} (2010) 385.
%
\bibitem{bra4} M.A.Braun, Eur.Phys. J. {\bf 66} (2010) 147.
%

\end{thebibliography}
\end{document}